\documentclass[11pt,a4paper]{article}

\usepackage[T1]{fontenc}
\usepackage[utf8]{inputenc}
\usepackage[english]{babel}

\usepackage{lmodern}
\usepackage{microtype}
\usepackage{geometry}
\usepackage[hidelinks]{hyperref}

\title{Causality in Physics:\\
From Galileo to Einstein, and Beyond}
\author{Alessandro De Angelis\\
University of Padua}
\date{Florence, September 2024}

\begin{document}
\maketitle

\paragraph{}
Causality is one of the most fundamental---and yet elusive---concepts in physics.
From its intuitive role in everyday experience to its formal and often implicit role in scientific theories,
causality has challenged philosophers and physicists alike.
In what follows, we take a brief historical and conceptual journey through classical and modern physics,
tracing how causality has been treated, questioned, or protected in successive physical frameworks---from Galilean mechanics to Newtonian dynamics,
from Lagrangian and Hamiltonian formulations to special and general relativity,
and finally to quantum mechanics and statistical physics.
Our aim is to show how the notion of causality has repeatedly receded into the background of our most successful theories,
even when it appears to be central to our everyday understanding of the world.

\section{The Birth of Mechanics: Galilei, Newton and the Question of Cause}

Causality---the principle that causes precede effects---seems at first glance a cornerstone of physical understanding.
Yet, somewhat surprisingly, physics lacks a general, formal definition of causality.
While causal language permeates the practice and communication of science,
the mathematical formulation of physical laws has largely operated without it.
This is a paradox: causality is both essential in our reasoning and undefined in our equations.

Galileo Galilei's \textit{Two New Sciences} (1638) represents a pivotal moment in the development of mechanics.
The four-day dialogue between the fictionalized interlocutors Simplicio, Sagredo, and Salviati explores two ``new sciences'':
the strength of materials and the motion of bodies.
The book provides an implicit notion of causality.
Though not framed in modern terms, Galileo introduces the principle of inertia,
which describes what happens in the absence of an external cause and underlies the later Newtonian understanding of motion.

In Galileo's dynamics, objects in motion continue their state unless acted upon by an external influence---an early recognition of what we might now describe as the cause of deviation from uniform motion (including rest).
Galileo sought to define how inertia is changed---essentially, what constitutes a ``cause'' in physics---but failed to complete this analysis before publication.
His intended ``Fifth Day,'' dealing with the \textit{Forza della percossa} (Force of percussion), was published only posthumously.
This unfinished quest highlights the conceptual difficulty of defining causality in mechanical terms.

Where Galileo laid the foundation, Isaac Newton built the structure.
In his \textit{Philosophiae Naturalis Principia Mathematica} (1687), Newton presents the three laws of motion.
The first, a formalization of the principle of inertia, sets the stage for a universe where change in motion must be accounted for by external influences.
The second law, $F=ma$, is commonly read as saying that forces cause accelerations:
the cause (force) results in an effect (change of motion).
Newton's absolute space and time provide a stable backdrop where the temporal order between cause and effect seems unambiguous.
Even here, however, things are subtler than they first appear.
Newton's second law is an equality: one can also write it as ``forces are what we ascribe to observed accelerations'',
particularly when working in accelerated frames of reference.
In such frames, so-called inertial forces appear---centrifugal force in a rotating frame, for example.
These ``forces'' arise from the acceleration of the frame itself rather than from any physical interaction.
In this sense, acceleration can be seen as the ``cause'' of such apparent forces.

This ambiguity becomes vivid in Newton's famous rotating bucket thought experiment.
The curved surface of the water in a rotating bucket, which takes on a parabolic shape, was taken by Newton as evidence for absolute rotation with respect to a fixed background (or at least to a privileged class of inertial frames).
In the 19th century, Ernst Mach questioned this conclusion,
suggesting instead that the origin of inertial effects is tied to the distribution of matter in the universe as a whole:
the bucket reacts not to empty space, but to all the masses in the cosmos.

Newton's third law---to every force, an equal and opposite reaction corresponds---adds further complexity.
It suggests a symmetry of interactions that does not by itself pick out a preferred direction in time.
Already in classical mechanics, the intuitive idea that ``cause comes before effect'' is not straightforwardly encoded in the equations.

\section{From Forces to Principles: Lagrangian and Hamiltonian Mechanics}

The 18th and 19th centuries saw a major reformulation of mechanics through the works of Joseph-Louis Lagrange and William Rowan Hamilton.
Instead of focusing on forces, Lagrangian mechanics describes systems in terms of generalized coordinates and a single function, the Lagrangian, typically equal to kinetic minus potential energy.
Hamiltonian mechanics similarly uses the Hamiltonian function, mostly representing the total energy of the system.

In this framework, the equations of motion arise from the principle of least (or more precisely, stationary) action.
Rather than writing ``force equals mass times acceleration'',
one demands that the integral of the Lagrangian along the actual path taken by the system be stationary with respect to small variations of that path.
Mathematically, this yields a set of differential equations---the Euler--Lagrange equations---that are equivalent to Newton's laws.

From the point of view of causality, this shift is striking.
The fundamental object is no longer a force that ``pushes'' objects around,
but a global variational principle that, at first sight, looks almost teleological:
it is as if the particle ``knows'' all the possible paths it could take and chooses the one that extremizes the action.
This impression has puzzled generations of students and has motivated philosophical reflection:
where has the familiar causal language gone?

At a deeper level, however, the Lagrangian and Hamiltonian formalisms do not abandon the differential-equation view of dynamics.
Time evolution is still local in time:
given the state of a system and its environment at one moment, the equations determine its state at later times.
What changes is not the underlying determinism, but the way we encode it mathematically.
Forces appear less as fundamental causes and more as derived quantities,
convenient for describing interactions but not the primary building blocks of the theory.

Modern physics has pushed this idea even further.
Symmetries of the Lagrangian or in the Hamiltonian---such as invariance under shifts in time or space---lead, via Noether's theorem, to conservation laws (of energy, momentum, and so on).
In this perspective, what is fundamental is not ``what causes what'',
but the symmetry structure of the laws themselves.

\section{Fields, Locality, and the Recasting of Influence}

The introduction of field theory in the 19th century, by Faraday and Maxwell, added a new layer to the causality puzzle.
A field assigns a value (scalar, vector, or tensor for example) to every point in space and time.
In electromagnetism, Maxwell's equations describe how electric and magnetic fields evolve and how they are related to their sources (charges and currents). Although we usually think densities of charge and current to cause the fiels, there is no such relation in Maxwell's equations.

A crucial step here is the elimination of instantaneous action at a distance.
Instead of charges exerting forces on each other across empty space,
changes in the electromagnetic field propagate at a finite speed (the speed of light).
The electromagnetic field becomes the mediator of influence.
The Lorentz force law describes how charged particles move in response to these fields,
while the fields themselves obey local differential equations, Maxwell's equations.

In this framework, causality becomes closely tied to locality and to the structure of the differential equations.
Given the field configuration and sources at an initial time, Maxwell's equations determine their evolution.
One can impose ``retarded'' boundary conditions, in which fields at a point depend only on sources in their past, not their future.
But again, this is a choice of physically reasonable solutions rather than something that is built into the raw mathematics of the equations.

The introduction of fields thus softens the old debates about action at a distance,
but it does not provide a clean, universal definition of causation.
Instead, it offers a more local and physically plausible mechanism for the propagation of influences.

\section{Mach, Russell, and the Critique of Causality}

At the dawn of the 20th century, Ernst Mach and Bertrand Russell articulated philosophical concerns about causality in physics.
Mach questioned the status of absolute space and time and emphasized the relational character of mechanics, foreshadowing later developments in relativity.
Russell, in his 1912 paper ``On the Notion of Cause'',
went even further, famously arguing that the concept of cause had no respectable role in fundamental physics and should be eliminated from scientific discourse.

For Russell, physical laws are best understood as systems of differential equations relating whole states of the world at different times,
rather than as lists of causes and effects.
The equations express patterns and correlations, not one-way arrows of influence.

He suggested, roughly, that if an event $e_1$ is always followed by an event $e_2$ after a certain time interval, one might call $e_1$ the cause of $e_2$.
But even in ordinary life this approach is unsatisfactory.
A classic example is the relation between night and day.
Night does not cause day, nor does day cause night.
Rather, both are correlated effects of a deeper mechanism: the rotation of the Earth.
Confusing correlation with causation is not just a philosophical faux pas; it is a recurring feature of human reasoning.

Russell's worries become even sharper when we consider that the fundamental laws of classical physics---Newton's laws, Maxwell's equations,
and even the basic equations of quantum mechanics---are, to a very good approximation, time-reversal symmetric.
They do not, by themselves, distinguish between past and future.
If one runs the equations backward, they still make sense.
In such a world, where does the arrow of ``cause then effect'' come from?

\section{Einstein's Relativity and Causal Structure}

Einstein's Special Relativity (1905) redefined space and time, merging them into a single four-dimensional spacetime.
In this picture, simultaneity is relative: whether two events happen ``at the same time'' depends on the observer's state of motion.
The old Newtonian background of absolute time disappears.

Relativity, however, does not abandon the idea of causal structure.
Quite the opposite: it makes this structure more explicit.
The limitation that no signal can travel faster than the speed of light gives rise to the notion of the light cone.
The future light cone of an event encompasses all points that can be reached by signals traveling at or below the speed of light;
the past light cone includes all points that can influence the event.
Events outside these cones cannot affect or be affected by it, in any reference frame.

This yields a relativistic definition of causal connection:
two events are causally related only if a signal traveling at speed $c$ or less can link them.
Any ``cause'' must lie within the past light cone of its ``effect''.
The structure of spacetime itself encodes which events can, in principle, influence which others.

General Relativity (1915) extends this framework to include gravitation, describing it as the curvature of spacetime produced by mass-energy.
Locally, the light-cone structure is preserved, and with it a local notion of causal order.
At the same time, the theory allows for exotic solutions---such as rotating universes or wormholes---with closed timelike curves,
which would in principle permit time travel and generate causal paradoxes.
Whether such solutions are physically realized is an open question,
and many physicists conjecture that ``chronology protection'' mechanisms forbid them in realistic situations
(this being often called ``Hawking chronology protection conjecture'').

Importantly, General Relativity also offers a new perspective on Mach's ideas and on inertial forces.
The equivalence between inertial and gravitational mass---central to Einstein's reasoning---blurs the distinction between gravitational and inertial forces.
In this picture, phenomena like the curvature of the water surface in Newton's rotating bucket could be consequences of spacetime geometry influenced by the distribution and motion of mass-energy (a modern incarnation of ``frame dragging'')
rather than signs of motion with respect to absolute space,
though the extent to which inertia is fully determined by cosmic matter (``Mach's principle'') remains debated.

From Galileo to Einstein, causality thus shifts from being tied to forces and absolute time
to being encoded in the geometric structure of spacetime, constrained by the finite speed of signal propagation.

\section{Quantum Mechanics, Path Integrals, and the Limits of Classical Causality}

Quantum theory adds a final, dramatic twist to the story.
In standard (``Copenhagen'') interpretations, a quantum system evolves deterministically according to the Schr\"odinger equation,
but measurement leads to an apparently instantaneous and probabilistic ``collapse'' of the wavefunction.
This collapse seems to introduce genuine indeterminism: even with complete knowledge of the state, we can predict only probabilities, not definite outcomes.

Entanglement makes the story even more puzzling.
Two particles that have interacted can display correlations that cannot be explained by any local hidden variables.
Measurements on one particle are correlated with measurements on the other, no matter how far apart they are.
Yet, as far as we know, these correlations cannot be used to send signals faster than light,
so relativistic causal structure---light cones and the prohibition of superluminal communication---remains intact.

The principle of least action reappears in quantum mechanics in an unexpected way through Feynman's path-integral formulation.
Instead of selecting a single path that extremizes the action, quantum mechanics assigns a complex amplitude to every possible path connecting two events in spacetime.
One then sums (integrates) over all these paths, and the squared modulus of the resulting amplitude gives the probability for the process.
Classical motion emerges, in a suitable limit, when the contributions from paths near the path of stationary action interfere constructively,
while those far away largely cancel.

This picture eliminates the apparent teleology of the classical least-action principle: the particle does not ``know'' the future.
Rather, all possible histories contribute to the amplitude, and interference among them selects the observed behavior.
The same formalism explains phenomena such as quantum tunneling (crucial, for example, in nuclear fusion processes in stars) and optical interference,
as Feynman famously illustrated.

Quantum mechanics also invites reinterpretations of the measurement process itself.
In ``many-worlds'' (Everett) interpretations, there is no physical collapse;
instead, measurement correlates (or entangles) the system with the measuring device and, ultimately, with the observer.
From the perspective of a single observer, it appears as if one outcome has been selected and others have vanished,
but at the fundamental level the dynamics is still deterministic.
Other approaches retain a genuine collapse but treat it as an additional stochastic law.

In all these interpretations, simple classical pictures of ``A causes B'' are hard to maintain at the microscopic level.
What the theory reliably delivers are probabilistic correlations constrained by symmetries, conservation laws, and the causal structure of spacetime.

\section{Irreversibility, Thermodynamics, and Emergent Causality}

Up to this point, the story may seem to support Russell's pessimism:
fundamental physics deals in differential equations and correlations, not in one-way arrows of cause and effect.
Yet our everyday experience is full of irreversible processes:
a cup shatters but does not spontaneously reassemble; heat flows from hot to cold, not the reverse.
Where does this macroscopic arrow of time come from, and what does it have to do with causality?

The key lies in thermodynamics and statistical mechanics.
While the microscopic laws are (to an excellent approximation) time-reversal symmetric,
macroscopic systems are not: they obey the second law of thermodynamics, according to which entropy tends to increase in closed systems.
This increase is not absolute---it depends on initial conditions---but in our universe, with its low-entropy past, it provides a robust temporal orientation.

In out-of-equilibrium statistical mechanics, the connection between irreversibility and causality becomes more concrete.
When we study how a system responds to disturbances---say, how a material responds to an applied electric field---we typically assume that the response cannot precede the disturbance.
This assumption of ``no effect before the cause'' has strong mathematical consequences.
For example, in linear response theory, it leads to relations which tightly connect the dissipative and reactive parts of a system's response functions.
Here, causality is not a vague philosophical notion but a precise condition on how systems behave in time.

In this sense, causality can be seen as emergent from the combination of time-symmetric microscopic laws,
special initial conditions (a low-entropy past), and the coarse-grained description of macroscopic, out-of-equilibrium systems.
It is at this level that the intuitive idea that ``the effect comes after the cause'' is most clearly realized and experimentally testable.

\section{The Human Perspective and the Perception of Causality}

There is one more layer to this story: us.
Even if fundamental physics does not contain an explicit, universal notion of causality,
human beings (and other animals) live in an out-of-equilibrium world filled with reliable regularities and irreversible processes.
Over evolutionary time, our brains have been shaped to exploit these regularities---to detect patterns, to predict what will happen next, and to act effectively on the world.

This biological and cognitive perspective helps explain why we are so strongly attracted to causal narratives.
We tend to see causal connections everywhere, sometimes correctly (as when we infer that flipping a switch turns on a light)
and sometimes not (as in superstition or the attribution of agency where none exists).
Our everyday talk of causes and effects is deeply tied to the macroscopic, thermodynamic arrow of time that we inhabit,
not to the underlying microphysics.

From this viewpoint, causality is partly a feature of the models we use to make sense of the world.
It is how we compress and organize the torrent of correlations we observe into manageable stories:
this happened because that happened.
Physics, by contrast, often works just as well---or better---by focusing on equations, symmetries, and statistical patterns.

What, then, is the status of causality in physics?
Our historical tour suggests a somewhat paradoxical conclusion.
Even though causal talk is central to how we, as human beings, think about the world,
causality has played a surprisingly modest role in the development of our most successful physical theories.

From Galileo and Newton onward, what has driven progress is not a sharpened definition of cause,
but the search for precise mathematical laws: differential equations, variational principles, symmetry structures, and probabilistic rules.
Forces, fields, and actions have been introduced and reinterpreted as our understanding has deepened,
but rarely with the explicit goal of defining causality.

Relativity shows that spacetime has a built-in structure of possible influences,
encoded in light cones and constrained by the speed of light.
Quantum mechanics and field theory reveal that, at the microscopic level,
we are often dealing with amplitudes and probabilities rather than clear-cut causal chains.
Thermodynamics and statistical mechanics show how an arrow of time---and with it a robust, experimentally meaningful notion of ``cause before effect''---can emerge in macroscopic, out-of-equilibrium systems.

Perhaps, then, causality is not a primitive element of reality, but an emergent, contextual concept.
At the fundamental level, physics provides us with correlations and constraints;
at the macroscopic level, in a universe with a low-entropy past,
those correlations can be organized into narratives of cause and effect.
And at the cognitive level, our brains exploit and sometimes overextend these narratives to navigate a complex world.

As we probe deeper into the structure of the universe,
the hope remains that future theories may clarify further how causal structure arises from more basic ingredients.
Until then, causality in physics is thought to be indispensable in practice but it is elusive in principle:
a concept we think we cannot do without, but one that our best equations are surprisingly reluctant to describe.

Causality is more a working hypothesis than an experimental fact.

\section*{References}

\begin{enumerate}
  \item Galileo Galilei, \textit{Discorsi e Dimostrazioni Matematiche intorno a Due Nuove Scienze}, Leiden, 1638. See also Alessandro De Angelis, \textit{Galileo Galilei's ``Two New Sciences'' for Modern Readers}, Springer, 2021.
  \item Isaac Newton, \textit{Philosophiae Naturalis Principia Mathematica}, 1687.
  \item J.\ L.\ Lagrange, \textit{M\'ecanique Analytique}, 1790.
  \item W.\ R.\ Hamilton, ``On a General Method in Dynamics,'' \textit{Philosophical Transactions of the Royal Society of London}, 1834--1835.
  \item J.\ C.\ Maxwell, ``A Dynamical Theory of the Electromagnetic Field,'' \textit{Philosophical Transactions of the Royal Society of London} 155 (1865), 459--512.
  \item Ernst Mach, \textit{The Science of Mechanics}, 1883.
  \item Bertrand Russell, ``On the Notion of Cause,'' \textit{Proceedings of the Aristotelian Society} 13 (1912--1913), 1--26.
  \item Albert Einstein, ``Zur Elektrodynamik bewegter K\"orper,'' \textit{Annalen der Physik} 17 (1905), 891--921.
  \item Albert Einstein, ``Die Feldgleichungen der Gravitation,'' \textit{Sitzungsberichte der Preussischen Akademie der Wissenschaften} (1915), 844--847.
  \item Roland Omn\`es, \textit{The Interpretation of Quantum Mechanics}, Princeton University Press, 1994.
  \item Ludwig Boltzmann, \textit{Lectures on Gas Theory} (1896--1898), translated by S.\ G.\ Brush, Dover, 1995.
  \item Ryogo Kubo, ``Statistical-Mechanical Theory of Irreversible Processes.\ I,'' \textit{Journal of the Physical Society of Japan} 12 (1957), 570--586.
  \item Richard P.\ Feynman, ``Space--Time Approach to Non-Relativistic Quantum Mechanics,'' \textit{Reviews of Modern Physics} 20 (1948), 367--387.
  \item Richard P.\ Feynman, \textit{QED: The Strange Theory of Light and Matter}, Princeton University Press, 1985.
  \item Hugh Everett III, ```Relative State' Formulation of Quantum Mechanics,'' \textit{Reviews of Modern Physics} 29 (1957), 454--462.
  \item John S.\ Bell, ``On the Einstein Podolsky Rosen Paradox,'' \textit{Physics} 1 (1964), 195--200; reprinted in \textit{Speakable and Unspeakable in Quantum Mechanics}, Cambridge University Press, 1987.
  \item H.\ Dieter Zeh, \textit{The Physical Basis of the Direction of Time}, 5th ed., Springer, 2007.
  \item Judea Pearl, \textit{Causality: Models, Reasoning, and Inference}, 2nd ed., Cambridge University Press, 2009.
  \item Albert Michotte, \textit{La Perception de la Causalit\'e}, Publications Universitaires de Louvain, 1946 (English translation: \textit{The Perception of Causality}, Basic Books, 1963).
  \item Huw Price, \textit{Time's Arrow and Archimedes' Point}, Oxford University Press, 1996.
\end{enumerate}

\end{document}